\documentclass[aps,prl,reprint,nobibnotes,
longbibliography,superscriptaddress]{revtex4-1}
\usepackage{t1enc}
\usepackage[utf8]{inputenc}
\usepackage{amsmath}
\usepackage{graphicx}
\usepackage{siunitx}
\usepackage{textcomp}

\begin{document}

\title{Indium as a high cooling power nuclear refrigerant for quantum
nanoelectronics}

\author{Nikolai Yurttagül}
\thanks{These authors contributed equally to this work.}
\author{Matthew Sarsby}
\thanks{These authors contributed equally to this work.}

\author{Attila Geresdi}
\email[Corresponding author; e-mail address: ]{a.geresdi@tudelft.nl}

\affiliation{QuTech and Kavli Institute of Nanoscience, Delft University of
Technology, 2600 GA Delft, The Netherlands}

\begin{abstract}
The frontiers of quantum electronics have been linked to the discovery of new
refrigeration methods since the discovery of superconductivity at a temperature
around $4\,$K, enabled by the liquefaction of helium. Since then, the advances
in cryogenics led to discoveries such as the quantum Hall effect and new
technologies like superconducting and semiconductor quantum bits. Presently,
nanoelectronic devices typically reach electron temperatures around $10\,$mK to
$100\,$mK by commercially available dilution refrigerators. However, cooling
electrons via the encompassing lattice vibrations, or phonons, becomes
inefficient at low temperatures. Further progress towards lower temperatures
requires new cooling methods for electrons on the nanoscale, such as direct
cooling with nuclear spins, which themselves can be brought to microkelvin
temperatures by adiabatic demagnetization. Here, we introduce indium as a
nuclear refrigerant for nanoelectronics and demonstrate that solely on-chip
cooling of electrons is possible down to $3.2\pm0.1\,$mK, limited by
the heat leak via the electrical connections of the device.
\end{abstract}

\maketitle

Quantum electronics relies on the precise control of
electronic states in nanostructures, which is possible if the energy-level
separation is much higher than the thermal energy $k_\textrm{B} T$. 
Access to novel states of matter such as electron-nuclear ferromagnets
\cite{deshpande2010,brauneckerhelical,chekhovich2013}, non-Abelian anyons in
fractional quantum Hall states \cite{Nayak_2008, stern2010}, topological
insulators \cite{RevModPhys.82.3045} or exotic superconductivity
\cite{RevModPhys.75.657, 0034-4885-66-12-R01,Klinovaja1} requires further
progress in the cooling of nanoelectronics, approaching the microkelvin regime.

Typical electron temperatures on the order of $10\,$mK are accessible in
semiconductor and metallic nanostructures by mounting the chip containing the
devices on an insulating substrate cooled by commercially available dilution
refrigerators. The lowest achievable electron temperature is limited by the heat
transferred from the electrons at a temperature of $T_e$ to phonons at a
temperature of $T_p$. The heat flow between conduction electrons and phonons in
a metallic volume $V$ is $\dot{Q}_\textrm{ep}=\Sigma V\left(T_e^5-T_p^5\right)$,
where $\Sigma$ is a material-dependent coupling constant on the order of
$10^9\,$WK$^{-5}$m$^{-3}$ \cite{RevModPhys.78.217, hotelectron}.
A residual electronic heat leak of $\dot{Q}_\textrm{leak}=10\,$aW to a
well-shielded nanostructure \cite{PhysRevB.85.012504} with $V=$
\SI{1}{\micro\meter\cubed} then yields $T_e\approx25\,$mK even as $T_p$
approaches zero.
Increasing the coupling volume $V$ by electrodeposition of thick metal films
\cite{bradley2016nanoelectronic} and by improving thermalization by means of
liquid-helium immersion cells led to steady-state values of $T_e\approx 4\,$mK
\cite{XIA2000491, doi:10.1063/1.3586766, bradley2016nanoelectronic} in specially built dilution
refrigerators.

The key to reduce the electron temperature further thus involves coupling the
electron system to a cold bath without the necessity of heat transport via
phonons. This can be achieved by nuclear magnetic cooling \cite{palma2017and,
bradley2017chip}. In the limit of small Zeeman splitting compared with
$k_\textrm{B}T_n$, the magnetization of the nuclear spin system is $M\propto
B/T_n$ at a magnetic field of $B$ and a temperature of $T_n$. $T_n$ can be
reduced by adiabatically lowering the magnetic field from $B_i$ to $B_f$. In the
absence of an external heat load, $M$ stays constant, and consequently
$T_{n,f}=T_{n,i}\times B_f/B_i$ \cite{gorter1934, kurti1956nuclear, progress}.
This technique has become the workhorse of ultralow-temperature physics, with
the lowest attainable temperature of $T_n\sim 100\,$pK
\cite{PhysRevLett.85.2573}.

On-chip nuclear magnetic cooling utilizes the spin-lattice
relaxation to cool conduction electrons close to the temperature of a
cold nuclear spin system that is co-integrated with the electronic device. The
heat flow $\dot{Q}_\textrm{en}$ is determined by the spin-lattice relaxation time
$\tau_1$ \cite{ensslowtemp, pobellmatter}:
\begin{equation}
\dot{Q}_\textrm{en} = \tau_1^{-1}nC_n\left({T_n-T_n^2/T_e}\right),
\end{equation}
where $C_n$ is the nuclear heat capacity and $n$ is the molar amount of
the nuclei.
If the magnetization is weak, the Korringa law $\tau_1T_e=\kappa$ applies with the
Korringa constant $\kappa$, and $C_n$ can be approximated by the Schottky law
$C_n=\alpha B^2/T_n^2$, where $\alpha=N_0I(I+1)\mu_n^2g_n^2/3k_\text{B}$, with
$I$ being the size of the nuclear spin, $g_n$ the g factor, $N_0$ the
Avogadro number and $\mu_n$ the nuclear magneton. In this limit, Eq.~(1) reads
\begin{equation}
\dot{Q}_\textrm{en}=\alpha n \kappa^{-1} B^2\left(T_e/T_n-1\right).
\end{equation}
The choice of nuclear refrigerant for nanoelectronics is based on finding a
material with a large $C_n$, while keeping $\kappa$ small to have efficient
coupling to the electrons. Eq.~(2) shows that the material-dependent
figure of merit $\alpha/\kappa$ allows us to compare different materials. In
addition, the experimental implementation should allow for a large $n$ and 
$B_i^2/T_{n,i}^2$ ratio for efficient cooling and long cold time.

\begin{figure*}
\centering
\includegraphics[width=\textwidth]{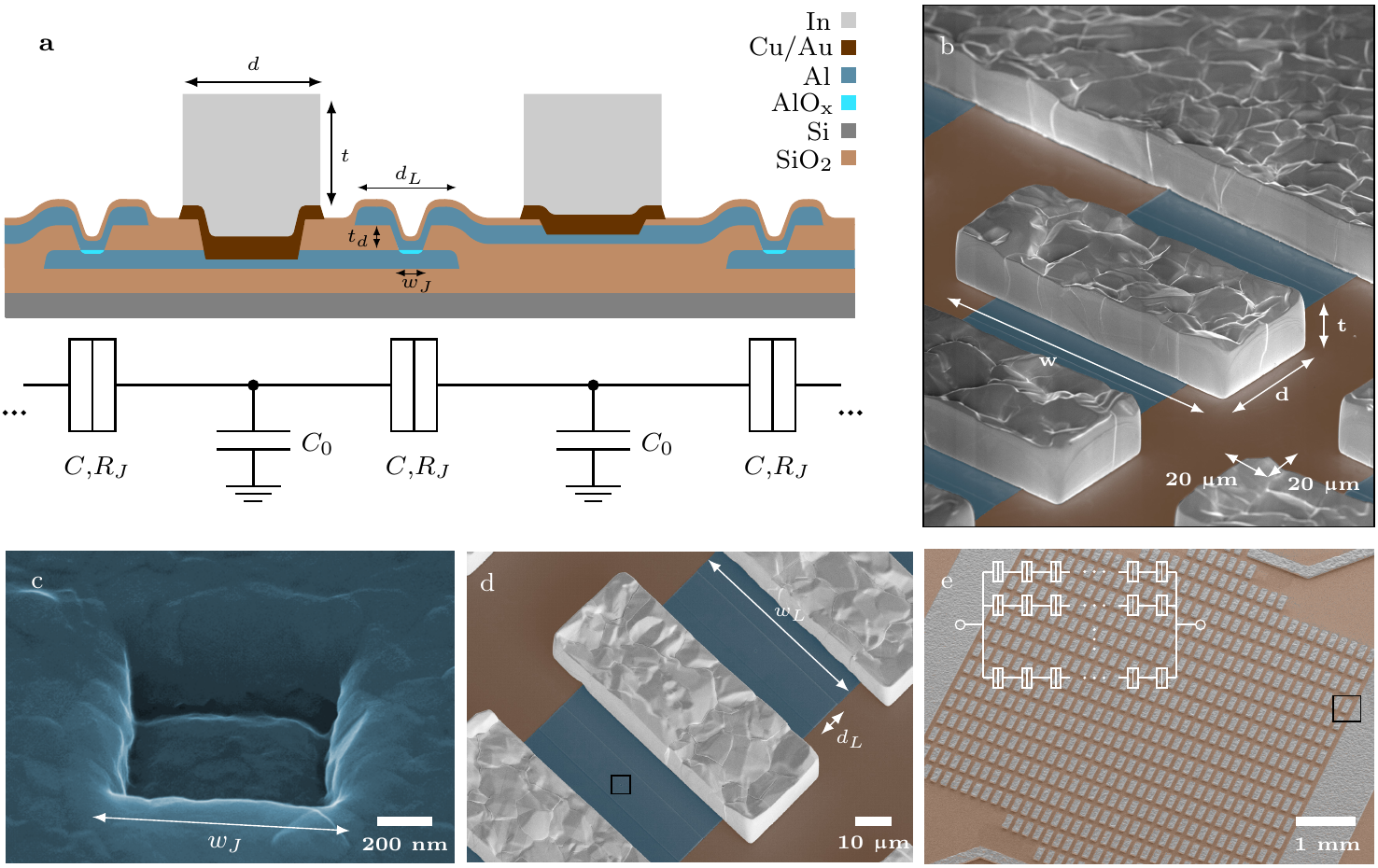}
\caption{(a) Cross section of the metallic islands
and the Al-AlO$_x$-Al tunnel junctions with 
a resistance of $R_J$. The capacitance $C$ is set by the overlap area $d_L
\times w_L$, $C_0$ is the stray capacitance. The electrons
are cooled by the electrodeposited In blocks.
(b) False-color scanning electron micrograph of a single In fin. (c) A single tunnel junction
between adjacent islands in a row (d). The black square in (d) depicts
the area shown in (c). (e) Overview of the full array with $35\times 15$
islands. The black square depicts the area shown in (b).}
\end{figure*}

Thus far, copper (Cu) was the sole material used for the nuclear cooling of
nanoelectronics \cite{doi:10.1063/1.3489892, palma2017and, bradley2017chip}.
Naturally occurring Cu nuclei have a spin of $3/2$, yielding
$\alpha_\textrm{Cu}=3.22\,$\si\micro JKT$^{-2}$mol$^{-1}$ \cite{pobellmatter}.
Bulk-Cu nuclear demagnetization stages benefit from the low magnetic ordering
temperature less than $0.1\,$\si\micro K \cite{symko1969nuclear}, which allows $T_n$
values in the microkelvin regime \cite{Xu1992}. However, the weak
electron-nucleus coupling given by $\kappa= 1.2\,$Ks \cite{ensslowtemp} is a
limitation for electron cooling.

To overcome this limit, we use indium (In) as an on-chip nuclear
refrigerant. In features a much shorter spin-lattice relaxation time
characterized by $\kappa_\textrm{In}=0.086\,$Ks \cite{korrindium}, and a large
nuclear spin of $9/2$ with $\alpha_\textrm{In}=13.8\,$\si\micro
JKT$^{-2}$mol$^{-1}$ increasing the molar nuclear cooling power by a factor of
$\frac{\alpha_\textrm{In}}{\kappa_\textrm{In}}/\frac{\alpha_\textrm{Cu}}{\kappa_\textrm{Cu}}=60$
compared with Cu. The lowest attainable $T_n$ for In is limited by
the tetragonal crystal-field splitting of $250\,$\si\micro K \cite{Tang1}. In
addition, the external magnetic field has to be kept above $B_c=28\,$mT to avoid
the thermal decoupling of electrons by the superconducting phase transition
\cite{Tinkham_1996}.

\begin{figure}
\includegraphics[width=0.5\textwidth]{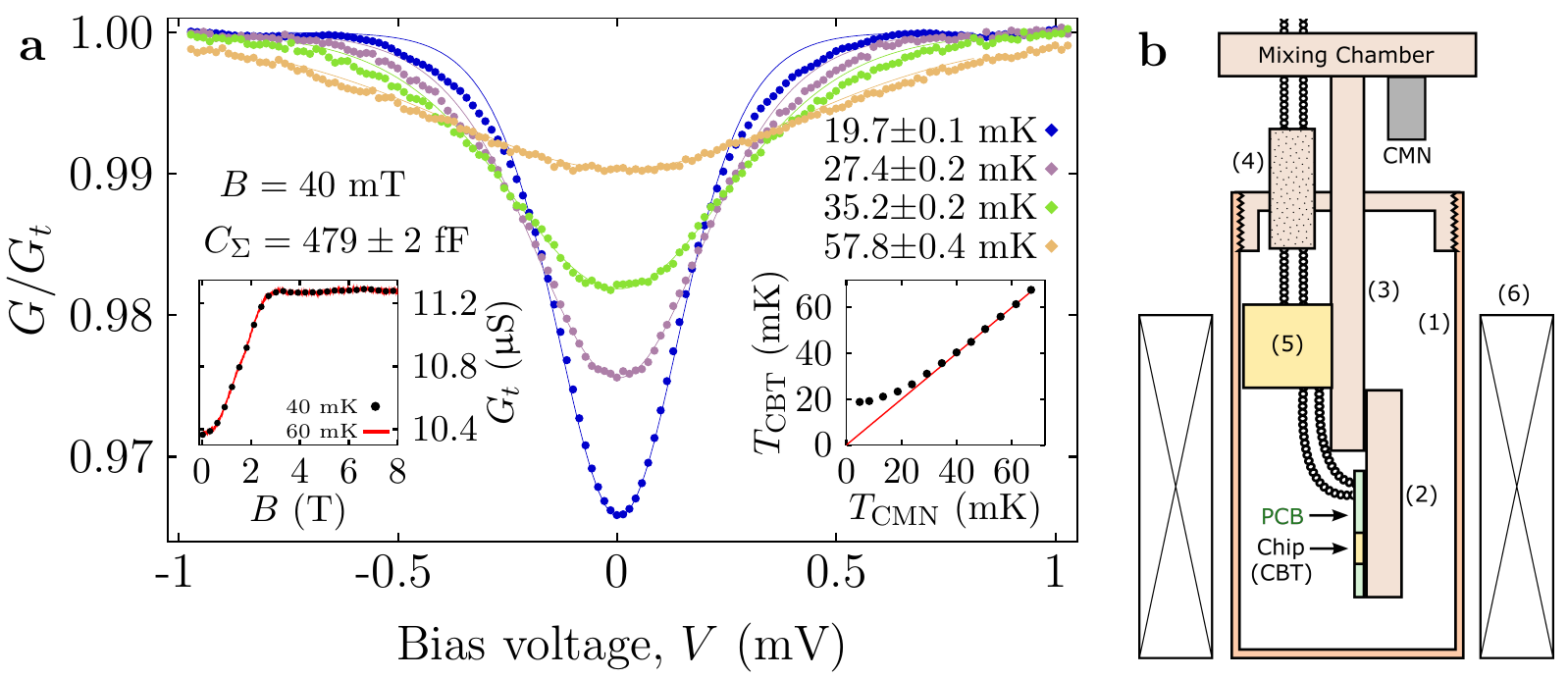}
\caption{(a) Normalized differential conductance $G/G_t$ as
a function of the voltage bias $V$ at different temperatures with the solid
lines showing the best fit determining the measured electron temperature
$T_\textrm{CBT}$ and the island capacitance $C_\Sigma=479\pm2\,$fF; see the text.
Right inset: $T_\text{CBT}$ as a function of the dilution refrigerator
temperature $T_\text{CMN}$ showing a saturated
$T_\text{CBT}\approx 20\,$mK. These measurements are taken with an applied magnetic
field $B=40\,$mT. Left inset: the change of
$G_t$ with magnetic field at two different temperatures.(b) The
cryogenic setup used in the experiment.
The CBT chip is mounted on a Cu plate (2), attached to the mixing chamber of the dilution refrigerator by a
Cu coldfinger (3).
The sample is well shielded from electromagnetic noise by an rf-tight enclosure
(1), Cu-powder filters (4) and resistive low-pass filters (5).
The magnetic field is applied by a superconducting solenoid (6).}
\end{figure}

To demonstrate the applicability of In as a refrigerant for scalable
nanoelectronics, we directly measure $T_e$ in a nanoelectronic device while
ramping $B$ to perform the nuclear demagnetization. While primary electron
thermometry in the millikelvin regime has been realized in several ways
\cite{Spietz1929, RevModPhys.78.217, feshchenko2015tunnel, iftikhar2016}, we
use Coulomb blockade thermometry due to its lack of sensitivity toa change in the magnetic
field \cite{doi:10.1063/1.367397}. Coulomb-blockade thermometers (CBTs) rely on the universal
temperature dependence of single charge localization in mesoscopic metallic
islands \cite{Pekola1} and have been proposed to provide the reference scale for
millikelvin-range thermometry \cite{meschke2011comparison}.

We integrate a nanofabricated CBT with In
cooling blocks to directly cool the electrons inside the device (Fig.~1).
The geometrical and electrical parameters of the thermometer are shown in
Fig.~1a.
The islands are formed within the stripe of $N=36$ tunnel junctions, and each
has a total capacitance $C_\Sigma=2C+C_0$, which determines its effective
charging energy $E_C=e^2/C_\Sigma \times (N-1)/N$, where $e$ is the elementary charge.
The zero-bias conductance of the device, $G(V=0)$ decreases by \cite{Pekola1, feshchenko2013primary}:
\begin{equation}
\frac{\Delta G}{G_t} = u_N/6 - u_N^2/60 + u_N^3/630 - \cdots,
\end{equation}
where  $u_N =  E_C/k_\textrm{B}T_e $. Notably, the width of the conductance dip
depends only on $N$ and $k_\textrm{B}T_e$, $eV_{1/2}/k_\text{B}T=5.44N$
enabling primary thermometry without prior calibration \cite{Farhangfar1}. We
set $E_C$ by the overlap area between adjacent islands, which is independent of
the tunnel junction area determining the junction resistance.

This flexibility in design is enabled by creating the Al-AlO$_x$-Al tunnel
junctions \emph{ex-situ} in vias through the interlayer dielectric (Fig.~1c)
\cite{prunnilaexsitu}.
The junction area is $w_j^2=0.55 \pm 0.1$~\si\micro m$^2$ with a tunnel resistivity of
$12.8\pm 0.8$~k$\Omega$\si\micro m$^2$ at room temperature, close to previously
reported values \cite{prunnilaexsitu}, yielding a total device resistance
$1/G_t=55.8\,$\si\kilo$\Omega$ in an array of $N\times M=36\times 15$ junctions.
The device resistance is temperature dependent and saturates at
$89\,$\si\kilo$\Omega$ below $T=1\,$K owing to the finite barrier height of the
AlO$_x$ insulating layer \cite{doi:10.1063/1.1320861}.

The CBT device used in this work is designed with an island overlap area of
$w_L\times d_L=18\times100\,$\si\micro m$^2$ and a SiO$_2$ interlayer dielectric
with thickness $t_d=230\,$nm (Fig.~1d). A parallel-plate-capacitor model
$C=\varepsilon_0\varepsilon_r A/d$ using $\varepsilon_r=3.5 - 3.9$ for sputtered
SiO$_2$ \cite{sioxdielectric} gives an estimated capacitance in the range of
$C_\Sigma=485 - 540\,$fF corresponding to the CBT working as a primary
thermometer between approximately $1.5$ and $250\,$mK. This range is limited by
uneven charge distribution on the low side \cite{feshchenko2013primary} and
instrumental resolution on the high side.

The nuclear-magnetic-cooling functionality is integrated by electroplating an In
block with a volume of
\SI{50}{\micro\metre}~$\times$~\SI{140}{\micro\metre}~$\times$~\SI{25.4\pm0.1}{\micro\metre}
through a thick photoresist mask onto each island (Fig.~1b). We achieve an In
integration density of 1.6~pmol/\si\micro m$^{2}$. This figure determines the
nuclear cooling power $\dot{Q}_\textrm{en}$ [see Eq.~(2)] per unit area.
Electroplating with a constant current results in extensive crystallization and
hydrogen formation which decreases the density of the films and affects the
patterning resolution. Therefore we apply forward-current and reverse-current pulses to
the electrochemical cell to refine the grain structure and reach a patterning
resolution of \SI{1}{\micro\metre}.

The device is mounted on a Cu carrier block inside an rf-tight enclosure
cooled by an unmodified commercial wet dilution refrigerator \cite{fridge}. The
mixing chamber temperature is measured by a calibrated cerium magnesium nitrate
(CMN) thermometer, and the base temperature is found to be approximately
$5\,$mK. The scheme of the measurement setup is shown in Fig.~2b. The CBT is
attached in a four-wire geometry with the twisted pairs of the electrical wiring
passing through a Cu powder \cite{pcbfilter} and a discrete-component third-order
RC low-pass filter, which has a cutoff frequency of $50\,$kHz to reduce external
noise. The differential conductance $G(V)$ of the CBT is measured by standard
low-frequency ($18.31\,$Hz) lock-in techniques as a function of the dc voltage
bias, $V$.

First we determine $E_C$ by simultaneously fitting a set of $G(V)$ curves
against the full single electron tunneling model \cite{Pekola1} at different
$T_\textrm{CMN}$ values set by heating the mixing chamber (Fig.~2a). We find 
$C_\Sigma=479\pm 2\,$fF, close to the designed value, yielding $E_C=330\,$neV.
The measured electron temperature $T_\textrm{CBT}$ agrees well with $T_\textrm{CMN}$ for
temperatures above $30\,$mK, but decouples and saturates for lower values,
demonstrating the inefficiency of phonon cooling via the device substrate (right
inset in Fig.~2a). The device exhibits a slightly-$B$-dependent
$G_t$ (left inset in Fig.~2a). This magnetoresistance is, however, independent of the device
temperature below $100\,$mK and therefore can be accounted for during the
demagnetization cycles. The established calibration of the CBT enables its use
in a secondary mode of operation, by measurement of the zero-bias conductance
decrease and finding $T_\textrm{CBT}$ on the basis of Eq.~(3). This mode avoids additional Joule heating at
finite bias voltages and allows for a real-time temperature sampling while the magnetic field is being ramped.

To perform the nuclear demagnetization experiment, we first set the initial
field $B_i$ and let the CBT thermalize while the heat released by nuclear spin
magnetization is absorbed by the dilution refrigerator.
We typically find $T_\textrm{CBT}\approx 20\,$mK after $24$ hours and $T_\textrm{CBT}\approx
16\,$mK after $72$ hours of precooling at $B_i=12.8\,$T. Then we reduce the
field to $B_f<B_i$ with a constant rate $\dot{B}$ while measuring the zero-bias
conductance $G(V=0)$ of the CBT.
We continuously track the conductance minimum in a dynamic bias window of
$\approx$~\SI{20}{\micro\volt} to compensate for voltage-bias drifts over the
several hours timescale of the experiment. The voltage-bias window
is continuously updated following the zero-bias point, which is determined by
the measurement software on the basis of the nonlinear $G(V)$ curve.

\begin{figure}
\includegraphics[width=0.5\textwidth]{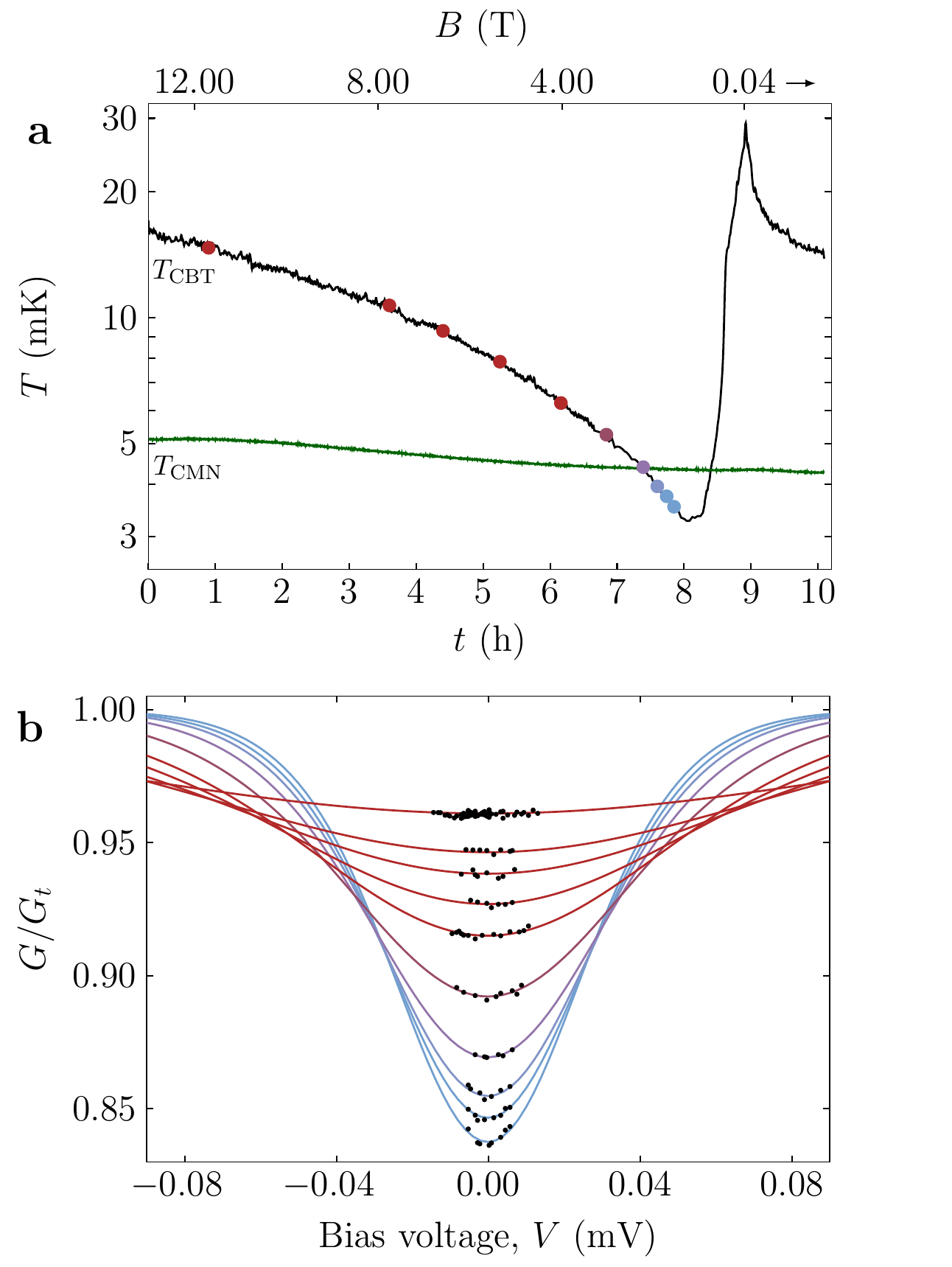}
\caption{(a) Electron
temperature $T_\textrm{CBT}$ and refrigerator temperature $T_\textrm{CMN}$ as
the field is reduced from $12.8\,$T (initial value $16\,$mK) to $40\,$mT at 
$\dot{B}=-0.4\,$ mT/s including a $1$-h thermal-relaxation period at the end.
$T_\textrm{CBT}<T_\textrm{CMN}$ is clearly obtained. (b)
The normalized conductance at small voltage biases (black dots) at different
temperatures confirms the expected zero-bias curvature; see the text. The position of each curve is shown by
a dot of the corresponding color in (a).}
\end{figure}

A typical experimental run starting from $B_i=12.8\,$T is shown in Fig.~3a.
We set a magnetic field ramp rate $\dot{B}=-0.4\,$mT/s and a final field
$B_f=40\,$mT. The strong electron-nucleus coupling results in a
reduction of the electron temperature $T_\textrm{CBT}$ from its initial value of $16\,$mK,
while $T_\textrm{CMN}\approx 5\,$mK remains essentially constant. During the
demagnetization cycle, we measure the curvature of the Coulomb-blockade peak and
find excellent agreement with the model calculations based on the already
inferred $E_C$ for each temperature data point with no additional fit parameters
(Fig.~3b). This analysis confirms the primary nature of our CBT
throughout the entire temperature range of the experiment.

Here we find an electron temperature of $T_e
=3.2\pm0.1\,$mK after on-chip demagnetization, with previously
reported values reaching $T_e=4.7\,$mK, with Cu as the nuclear
refrigerant \cite{bradley2017chip}. We also demonstrate superior performance
compared with phonon-cooled devices, where steady-state values above $T_e=3.9\,$mK
were measured in a custom-built dilution refrigerator
\cite{bradley2016nanoelectronic}.

On reaching the lowest temperature at $B\approx2$ T, the stage warms up
quickly, and after the field ramp stops at $B_f=40\,$mT, it relaxes to the
starting temperature. This behavior indicates that the nuclear heat capacity
is depleted by parasitic heat leaks before the field ramp is finished. We analyze
the cooling performance and heat leaks of our device by performing
demagnetization runs at different $\dot{B}$ between $-0.3\,$mT/s and $-6\,$mT/s
(Fig.~4a). All data exhibit a similar behavior with a minimum temperature
well below $T_\textrm{CMN}\approx 5\,$mK. To determine the heat input by
parasitic heating, we numerically model the time evolution of $T_n$ and $T_e$ on
a single island. We ignore the weak thermal coupling to phonons and assume that
the heat flow between the electrons and nuclei is described by Eq.~(1). We
consider the Hamiltonian of the nuclear spin where the
interaction of the nuclear quadrupole moment with the crystal-field gradient is
included with $e^2qQ=-198\,$neV in a direction $\Theta$ with respect to a large
external magnetic field $B$ \cite{Tang1}:
\begin{equation}
\mathbf{H} = -\gamma B\mathbf{I}_z +
\frac{e^2qQ}{4I(2I-1)}\frac{3\cos^2\Theta-1}{2}\left(3\mathbf{I}_z^2-\mathbf{I}^2\right).
\end{equation}
The set of eigenvalues $\varepsilon_m$ are averaged over $\Theta$ and
define the partition sum $Z=\sum \exp(-\varepsilon_m/k_B T_n)$, which
yields the nuclear spin entropy $S_n=k_\text{B}\partial(T_n\log Z)/\partial T_n$
and the heat capacity $C_n=T_n(\partial S_n/\partial T_n)_B$. The time evolution of the temperatures
then follows:
\begin{equation}
\begin{aligned}
\frac{dT_n}{dt}
&=\frac{T_eT_n-T_n^{2}}{\kappa}+\frac{\partial T_n}{\partial B}\dot{B}, \\
\frac{dT_e}{dt}
&=-\frac{C_n}{C_e}\left(\frac{T_eT_n-T_n^{2}}{\kappa}\right)+\frac{\dot{Q}_\textrm{leak}}{nC_e}.
\end{aligned}
\end{equation}
The heat capacity of the electron system is assumed to follow the Sommerfeld
rule, $C_e=\gamma T_e$, and we include a parasitic heat leak
$\dot{Q}_\textrm{leak}$ as a free parameter. Evaluating the experimental data in
Fig.~4a, we plot $Q_\textrm{leak}(B)$ in the upper inset in Fig.~4a and find the
same behavior independent of $\dot{B}$ with a linear increase at high fields
and a rapid upturn below $2\,$T. We evaluate the linear segment and find that
the dissipation $\dot{Q}_\text{leak}$ changes linearly with $\dot{B}$ (lower
inset in Fig.~4), $\dot{Q}_\text{leak}=a |\dot{B}|+\dot{Q}_0$, with $a
=18\,$fW/(mT~s$^{-1}$) and $\dot{Q}_0 = 0.18\,$fW per island. Notably, the
largest measured heat leak per island is $108\,$fW, similar to earlier
reported values in similar experiments \cite{bradley2017chip}, but much higher
than figures for combined on-chip and off-chip refrigeration \cite{palma2017and,
sarsby500muK}.

The linear $\dot{Q}_\textrm{leak}(\dot{B})$ is in striking contrast with
observations on macroscopically large nuclear cooling stages, where eddy currents in the bulk
refrigerant lead to $\dot{Q}_\textrm{leak} \propto \dot{B}^2$
\cite{pobellmatter}. It is also inconsistent with a dominating static heat leak
yielding a constant $\dot{Q}_\textrm{leak}$, or with an  environmental coupling
of $\dot{Q}_\textrm{leak} \propto T_e^n - T_\textrm{env}^n$.
This confirms that the metallic islands of the CBT are thermally well decoupled
from the environment; however, the rate-independent $Q_\textrm{leak}$ suggests
the presence of a well-coupled thermal mass consuming the In nuclear heat
capacity over the timescale of the experiment.

\begin{figure}
\includegraphics[width=0.5\textwidth]{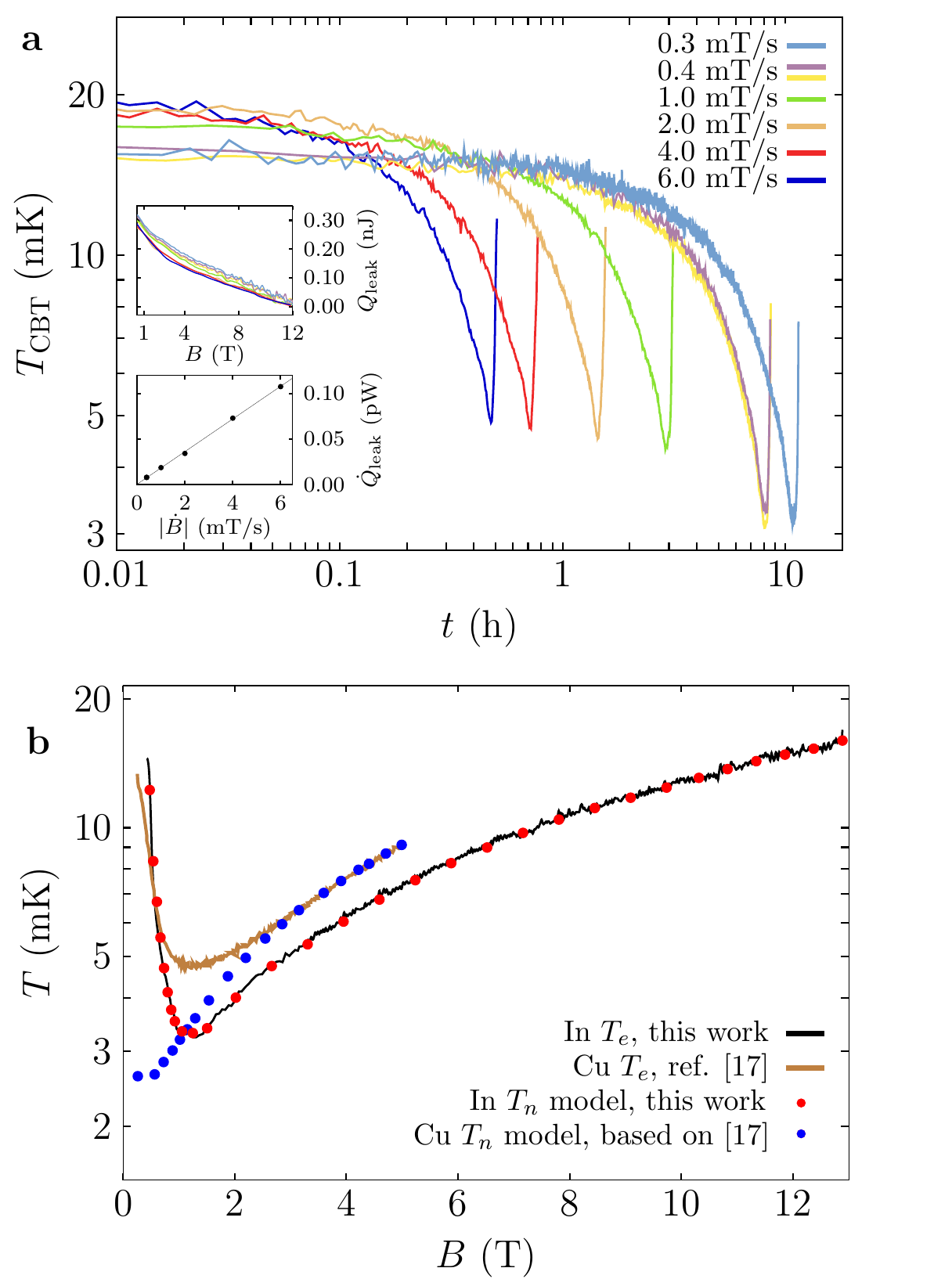}
\caption{(a) $T_\text{CBT}$ as a function of time for various magnetic
field ramp rates, $\dot B$. Two distinct runs with $\dot{B}=-0.4\,$mT/s are
shown to demonstrate the reproducibility of the data. The lower inset shows the
extracted heat leak $\dot{Q}_\textrm{leak}$ per island (black dots) and a linear
fit (solid line). The upper inset demonstrates that the total absorbed heat
$Q_\textrm{leak}$ per island collapses onto a single curve for all runs; see the
text. (b) Comparison between on-chip demagnetization performed in the work reported in Ref.~\cite{bradley2017chip} with Cu (raw data retrieved from \cite{rawdataLanc}) and
our experiment using In with $\dot{B}=-0.4\,$mT/s. The $T_e$ values
(solid lines) are measured data, whereas $T_n$ (dots) is calculated with Eq.~(5).}
\end{figure}

We evaluate the performance of electroplated In as an on-chip electron
refrigerant by considering the nuclear heat capacity integrated per unit
area, $\alpha^\prime = \alpha n / A$ and use $\alpha^\prime/\kappa$ as a figure
of merit. Our
device features $\alpha^\prime/\kappa_\textrm{In}=250\,$\si\micro
W/(m$^2$KT$^2$), to be compared with $0.076\,$\si\micro W/(m$^2$KT$^2$)
\cite{palma2017and} and $2.33\,$\si\micro
W/(m$^2$KT$^2$)  \cite{bradley2017chip}, achieved with Cu. 

We benchmark our implementation by comparing our data with the results 
reported in Ref.~\cite{bradley2017chip}. The measured $T_e$ and the calculated
$T_n$ are plotted in Fig.~4b. Our simulations based on Eq.~(5) confirm that the
In nuclei are strongly coupled to the electron system, resulting in $T_n=T_e$
throughout the entire experiment with $\dot{B}=-0.4\,$mT/s, resulting in
$\dot{Q}_\textrm{leak}=7.85\,$fW per island. In contrast, the weaker hyperfine
coupling of Cu results in a deviation between $T_n$ and $T_e$ despite a similar
heat leak $\dot{Q}_\textrm{leak}=6.3\,$fW per island \cite{bradley2017chip},
demonstrating the limitations of Cu as an on-chip refrigerant.

In conclusion, we demonstrate that electron temperatures of $3.2\pm0.1\,$mK can
be reached by nuclear magnetic cooling on-chip with In as nuclear refrigerant, a
result that is the coldest measured electron temperature without additional
off-chip nuclear demagnetization cooling. We therefore conclude that on-chip
integrated nuclear refrigeration using In is a versatile means to decrease the
electron temperature of nanoscale devices.

\begin{acknowledgments}
We thank J.~Mensingh and  R.~N.~Schouten for technical assistance. We
acknowledge O.~Benningshof, J.~Pekola and R.~Haley for fruitful discussions and
comments on the manuscript. This work was supported by the Netherlands
Organization for Scientific Research (NWO) and Microsoft Corporation Station Q.
\end{acknowledgments}

Raw datasets measured and analyzed for this publication are available at the
4TU Centre for Research Data repository \cite{rawdata}.

\bibliography{references_on_chip_demag}

\end{document}